\documentstyle[12pt,graphicx]{article}
\begin{document}
\title{Euclidean Scalar Green Function in a Higher Dimensional
Global Monopole Spacetime}
\author{E. R. Bezerra de Mello \thanks{E-mail: emello@fisica.ufpb.br}
\\
Departamento de F\'{\i}sica-CCEN\\
Universidade Federal da Para\'{\i}ba\\
58.059-970, J. Pessoa, PB\\
C. Postal 5.008\\
Brazil}
\maketitle
\begin{abstract}
We construct the explicit Euclidean scalar Green function associated with
a massless field in a higher dimensional global monopole spacetime, i.e.,
a $(1+d)$-spacetime with $d\geq3$ which presents a solid angle deficit. 
Our result is expressed in terms of a infinite sum of products of Legendre 
functions with Gegenbauer polynomials. Although this Green function cannot 
be expressed in a closed form, for the specific case where the solid angle 
deficit is very small, it is possible to develop the sum and obtain the 
Green function in a more workable expression. Having this expression it is 
possible to calculate the vacuum expectation value of some relevant operators.
As an application of this formalism, we calculate the 
renormalized vacuum expectation value of the square of the scalar field,
$\langle \Phi^2(x)\rangle_{Ren.}$, and the energy-momentum tensor, 
$\langle T_{\mu\nu}(x)\rangle_{Ren.}$, for the global monopole spacetime 
with spatial dimensions $d=4$ and $d=5$. 
\end{abstract}

\newpage
\renewcommand{\thesection}{\arabic{section}.}
\section{Introduction}
In this paper we consider the Euclidean scalar Green function associated 
with a massless field in the higher dimensional global monopole spacetime.
We define this spacetime as a generalization of the previous one given
in Ref. \cite{BV} (the metric of cosmic string spacetime which is spatially
flat was considered in Ref. \cite{Helliwell}). The generalization of the
Euclidean line element of \cite{BV} for higher dimensional case is given
by 
\begin{equation}
ds^2=d\tau^2+\frac{dr^2}{\alpha^2}+r^2d\Omega_{d-1}^2=g_{\mu\nu}(x)
dx^\mu dx^\nu \ ,
\end{equation}
where $\mu,\  \nu=0, 1, 2 ...    d$, with $d\geq3$ and $x^\mu=(\tau, r, 
\theta_1, \theta_2, ...  ,\theta_{d-2}, \phi)$. The coordinates are defined 
in the intervals $\tau\in(-\infty, \infty)$, $\theta_i\in [0, \pi]$ for 
$i=1, 2 ... d-2$, $\phi\in[0, 2\pi]$ and $r\geq0$. The parameter $\alpha$
which codify the presence of the global monopole is smaller than unity. 
In this coordinate system the metric tensor is explicitly defined as shown 
below:
\begin{eqnarray}
\label{Metric}
g_{00}&=&1 \nonumber \\ 
g_{11}&=&1/\alpha^2 \nonumber \\ 
g_{22}&=&r^2 \nonumber\\ 
g_{jj}&=&r^2\sin^2\theta_1\sin^2\theta_2 ...\sin^2\theta_{j-2} \ ,
\end{eqnarray}
for $3\leq j \leq d$, and $g_{\mu\nu}=0$ for $\mu\neq \nu$.

This spacetime corresponds to a pointlike global monopole. It is not
flat: the scalar curvature $R=(d-1)(d-2)(1-\alpha^2)/r^2$, and the solid 
angle associated with a hypersphere with unity radius is $\Omega=2\pi^{d/2}
\alpha^2/\Gamma(d/2)$, so smaller than ordinary one. The energy-momentum
tensor associated with this object has a diagonal form and its 
non-vanishing components read $T_0^0=T_1^1=(\alpha^2-1)(d-2)/r^2$.

The Euclidean scalar Green function associated with a massless field in
the geometry defined by (\ref{Metric}) should obey the non-homogeneous
second order differential equation
\begin{equation}
\left(\Box-\xi R\right)G_E(x,x')=-\delta^n(x,x')=-\frac{\delta^n(x-x')}{\sqrt{g}} \ ,
\end{equation}
where $n=d+1$, $\Box$ denotes the covariant d'Alembertian in the spacetime
defined by (\ref{Metric}), $\xi$ is an arbitrary coupling constant, 
$\delta^d(x,x')$ is the bidensity Dirac distribution and $R$ is the scalar
curvature.

The vacuum expectation value of the square of the scalar field, $\langle
\Phi^2(x)\rangle$, is given by the evaluation of the Green function at the 
same point which provides a divergent result. So, in order to obtain a finite
and well defined one, we should apply some renormalization procedure. The 
method which we shall adopt here is the point-splitting renormalization 
procedure. The basic idea of this method consists of subtracting from the 
Green function all the divergences
which appears in the coincidence limit. (As it is well known this procedure
is ambiguous in even dimensions and the ambiguity is a consequence of 
mass scale parameter $\mu$  which is introduced after the renormalization 
procedure. In this way the final result contains a finite part which 
depends on the scale $\mu$.) 

In \cite{Wald} Wald observed that the singular behavior of the Green function 
in the coincidence limit has the same structure as the Hadamard function, 
so his proposal was to subtract from the Green function the Hadamard one.
In fact the use of the Hadamard function to renormalize the vacuum expectation
value of the energy-momentum tensor in a curved spacetime has been first
introduced by Adler {\it et. al.} \cite{Adler}. In Ref. \cite{Wald} Wald
added a modification to this technique in order to provide the correct
result for the trace anomaly. In this way the renormalized vacuum expectation 
value of the square of the field operator is given by:
\begin{equation}
\langle\Phi^2(x)\rangle_{Ren.}=\lim_{x'\to x}\left[G_E(x,x')-
G_H(x,x')\right] \ .
\end{equation} 

The renormalized vacuum expectation value of the energy-momentum tensor
can also be obtained in a similar way:
\begin{equation}
\langle T_{\mu\nu}(x)\rangle_{Ren.}=\lim_{x' \to x}{\cal{D}}_{\mu\nu'}(x,x')
\left[G_E(x,x')-G_H(x,x')\right] \ ,
\end{equation}
where ${\cal{D}}_{\mu\nu'}(x,x')$ is a bivector differential operator. 
Moreover this vacuum expectation value should be conserved, i. e.,
\begin{equation}
\nabla_\mu\langle T^\mu_\nu(x)\rangle_{Ren.}=0 \ ,
\end{equation}
and gives us the correct conformal trace anomaly,
\begin{equation}
\langle T^\mu_\mu(x)\rangle_{Ren.}=\frac1{(4\pi)^{n/2}}a_{n/2}(x) \ ,
\end{equation}
for $n=1+d$ even and
\begin{equation}
\langle T^\mu_\mu(x)\rangle_{Ren.}=0 \ ,
\end{equation}
for $n=1+d$ odd \cite{Christensen}.

In this paper we study the quantum field theory associated with a 
massless scalar field in the background spacetime defined by (\ref{Metric}). 
More specifically we shall specialize in the cases where the spatial section
of our spacetime has dimensions $d=4$ and $d=5$.
In section $2$, we construct explicitly the Euclidean Green function 
for a massless scalar field in a higher dimensional global monopole spacetime. 
We show that this Green function is expressed in terms of an infinite sum
of product of  Legendre function with Gegenbauer polynomials. In section $3$ 
we calculate explicitly the renormalized vacuum expectation value of the 
square of the scalar field operator for $d=4$ and $d=5$ dimensions in the 
case where $\eta^2 \ll 1$, being $\eta^2=1-\alpha^2$\ \footnote{The parameter 
$\alpha$ is related with the energy scale where the global symmetry of the 
system is spontaneously broken, see Ref. \cite{BV}. For a typical grand 
unified theory in a physical $(1+3)-$dimensional spacetime, this scale 
is of order $10^{16}$Gev. So $1-\alpha^2=\eta^2 \sim 10^{-5}$.}. We show 
that for six dimensions the expression obtained presents an ambiguity given 
by the mass scale parameter and that this ambiguity vanishes if we assume for
the non-minimal coupling constant its conformal value, $\xi=1/5$. 
In section $4$ we present the formal expressions for the vacuum expectation 
values of the energy-momentum tensor for the cases $d=4$ and $d=5$. For the 
six dimensional case we present, after a long calculation, an explicit 
expression for the scale dependent term up to the first order in $\eta^2$. 
We explicitly show that this term is traceless for the conformal coupling. 
In section $5$ we present our conclusions and some important remarks about
this paper. Although the formalism developed here is applied only when the
dimension of the space section is $d\geq3$, in the appendix we present an 
expression for the Green function in a limiting case $d\to2$.

\section{Green Function}

In this section we  calculate the Euclidean scalar Green function
associated with a massless field in the spacetime described by (\ref{Metric}).
This Green function  must obey the non-homogeneous second order
differential equation 
\begin{equation}
\left(\Box-\xi R\right)G_E(x,x')=-\delta^d(x,x')=-\frac{\delta^2(x-x')}
{\sqrt{g}} \ ,
\end{equation}
where we have introduced the non-minimal coupling of the scalar field
with the geometry. As we have already said, the scalar curvature
for this spacetime is $R=(d-1)(d-2)(1-\alpha^2)/r^2$.

The Euclidean Green function can also be obtained by the Schwinger-DeWitt
formalism as follows:
\begin{equation}
\label{Heat}
G_E(x,x')=\int^\infty_0 ds K(x,x';s) \ ,
\end{equation}
where the heat kernel, $K(x,x';s)$, can be expressed in terms of 
eigenfunctions of the operator $\Box-\xi R$ as follows:
\begin{equation}
\label{Heat-1}
K(x,x';s)=\sum_\sigma \Phi_\sigma(x)\Phi_\sigma^*(x')
\exp(-s\sigma^2) \ ,
\end{equation}
$\sigma^2$ being the corresponding positively defined eigenvalue. Writing
\begin{equation}
\left(\Box-\xi R\right)\Phi_\sigma(x)=\sigma^2 \Phi_\sigma(x) \ ,
\end{equation}
we obtain the complete set of normalized solutions of the above equation:
\begin{equation}
\Phi_\sigma(x)=\sqrt{\frac{\alpha p}{2\pi}}\frac1{r^{d/2-1}}e^{-i\omega\tau}
J_{\nu_n}(pr)Y(n, m_j; \phi, \theta_j) \ ,
\end{equation}
with
\begin{equation}
\sigma^2=\omega^2+\alpha^2p^2 \ ,
\end{equation}
$Y(n,m_j;\phi, \theta_j)$ being the hyperspherical harmonics of degree $n$ 
\cite{Y}, $J_\nu$ the Bessel function of order
\begin{equation}
\nu_n=\alpha^{-1}\sqrt{\left(n+(d-2)/2\right)^2+(d-1)(d-2)(1-\alpha^2)
(\xi-\overline{\xi})} \ ,
\end{equation} 
with $\overline{\xi}=\frac{d-2}{4(d-1)}$. So according to (\ref{Heat-1}) our 
heat kernel is given by
\begin{eqnarray}
\label{Heat1}
K(x,x';s)&=&\int^\infty_{-\infty}d\omega\int^\infty_0 dp\sum_{n,m_j}
\Phi_\sigma(x)\Phi^*_\sigma(x')e^{-s\sigma^2}
\nonumber\\
&=&\frac1{8\alpha\pi^{\frac{d+1}2}s^{3/2}}\frac1{(rr')^{d/2-1}}
\frac{\Gamma(d/2)}{d-2}e^{-\frac{\Delta\tau^2\alpha^2+r^2+r'^2}{4\alpha^2
s}}\times
\nonumber\\
&&\sum_{n=0}^\infty[2(n-1)+d]I_{\nu_n}\left(\frac{rr'}{2\alpha^2 s} \right)
C_n^{\frac{d-2}2}(\cos \gamma) \ ,
\end{eqnarray}
$I_\nu$ being the modified Bessel function, $C_n^\mu(x)$ the Gegenbauer 
polynomial of degree $n$ and order $\mu$ and $\gamma$ is the angle between 
two arbitrary directions. Our final expression (\ref{Heat1}) was 
obtained using the addition theorem for the hyperspherical harmonics 
\cite{Y} and the integral table \cite{Gradshteyn}.

Now we are in position to obtain the Euclidean Green function substituting
(\ref{Heat1}) into (\ref{Heat}). Our final result is:
\begin{eqnarray}
\label{Green}
G^{(d)}_E(x,x')&=&\frac1{4\pi^{d/2+1}}\frac1{(rr')^{\frac{d-1}2}}
\frac{\Gamma(d/2)}{d-2}\times
\nonumber\\
&&\sum_{n=0}^\infty[2(n-1)+d]Q_{\nu_n-1/2}(u)
C_n^{\frac{d-2}2}(\cos\gamma) \ ,
\end{eqnarray}
where
\begin{equation}
u=\frac{\alpha^2\Delta\tau^2+r^2+r'^2}{2rr'} \ 
\end{equation}
and $Q_\mu$ is the Legendre function. Unfortunately because the dependence
of the order of the Legendre functions on the parameter $\alpha$ is not 
a simple one, it is not possible to develop the summation and to obtain
a closed expression for the Green function even in the simpler case $d=3$
\cite{Mazzitelli}. However for $\xi=\overline{\xi}$, $\nu_n$ becomes equal to
$(n+(d-2)/2)/\alpha$ and for $\gamma=0$ it is possible to represent this Green 
function in an integral form using the integral representation for the 
Legendre function
\begin{equation}
\label{Legendre}
Q_{\nu-1/2}(\cosh\rho)=\frac1{\sqrt{2}}\int^\infty_\rho dt
\frac{e^{-\nu t}}{\sqrt{\cosh t-\cosh\rho}}
\end{equation}
and 
\begin{equation}
C_n^{\frac{d-2}2}(1)=\frac{(n+d-3)!}{n!(d-3)!} \ .
\end{equation}
Now it is possible to develop the summation and after
some intermediate steps we get
\begin{eqnarray}
G^{(d)}_E(r,\tau;r',\tau')&=&\frac1{4\pi^{d/2+1}}\frac1{(rr')^{\frac{d-1}2}}
\frac{\Gamma(d/2)}{2^{d-3/2}}\times
\nonumber\\
&&\int^\infty_\rho dt\frac1{\sqrt{\cosh t-\cosh\rho}}\frac{\cosh(t/2\alpha)}
{(\sinh(t/2\alpha))^{d-1}} \ .
\end{eqnarray}
For the case where $\alpha=1/2$, i.e., a large solid angle deficit, the
above expression can be written in a simpler form by performing the
integral. Making an appropriate transformation of variable, 
$t:=\mbox{arccosh}\left(\frac{u-1}q+1\right)$, we get:
\begin{eqnarray}
G^{(d)}_E(r, \tau; r, \tau')&=&\frac1{4\sqrt{2}\pi^{d/2+1}}\frac{\Gamma(d/2)}
{d-2}\frac{(rr')^{\frac{d-3}2}}{[(r-r')^2+\Delta\tau^2/4]^{d/2-1}}\times
\nonumber\\
&&\frac1{[(r+r')^2+\Delta\tau^2/4]^{d/2-1}} \ .
\end{eqnarray}
 
The next section is devoted to the obtainment of the renormalized 
vacuum expectation value of the square of the field operator. As an
explicit application we shall develop these quantities for the cases where
$d=4$ and $d=5$.

\section{Calculation of $\langle\Phi^2(x)\rangle_{Ren.}$}

The calculation of the vacuum expectation value of the square of the
field operator is obtained computing the Green function in the coincidence
limit:
\begin{equation}
\langle\Phi^2(x)\rangle=\lim_{x'\to x}G_E(x,x') \ .
\end{equation}
However this procedure provides a divergent result. In order to 
obtain a finite and well defined result we must apply in this calculation
some renormalization procedure. Here we shall adopt
the point-splitting renormalization. This procedure is based
upon a divergence subtraction scheme in the coincidence limit of the
Green function. In \cite{Wald}, Wald examined the behavior of the Green
function in this limit. He observed that its divergences have the same
structure as given by the Hadamard function, which on the other hand
can be explicitly written in terms of the square of the geodesic 
distance between the two points. So, we shall adopt the
following prescription: we subtract from the Green function the Hadamard
one before applying the coincidence limit as shown below:
\begin{equation}
\label{Phi2}
\langle\Phi^2(x)\rangle_{Ren.}=\lim_{x'\to x}\left[G_E(x,x')-
G_H(x,x')\right] \ .
\end{equation}

Now let us develop this calculation explicitly. As we have mentioned
above, it is not possible to proceed exactly the summation which appears
in the Green function. By this reason the best that we can do is to obtain 
an approximate
expression for it, developing a series expansion in powers of the parameter 
$\eta^2=1-\alpha^2$, which is much smaller than unity for the physical 
four-dimensional spacetime. Moreover, because we want to take the coincidence
limit in the Green function, let us take first $\gamma=0$ and $\Delta\tau=0$
in (\ref{Green}). The approximate expression for the order of the Legendre 
function, up to the first power in $\eta^2$ is
\begin{equation}
\nu_n\approx \left(n+\frac{d-2}2\right)(1+\eta^2/2)+\frac{(d-1)(d-2)(\xi-
\overline{\xi})}{2n+d-2}\eta^2+O(\eta^4) \ .
\end{equation}
We also need to develop the summation
\begin{equation}
S=\sum_{n=0}^\infty[2n+d-2]\frac{(n+d-3)!}{n!}e^{-\nu_nt} \ .
\end{equation}
Substituting the approximate expression for $\nu_n$ into the summation above,
we get after some intermediate steps,
\begin{equation}
S=\frac{(d-2)!}{2^{d-2}}\frac{\cosh(t/2)}{\sinh^{d-1}(t/2)}\left[1
-\frac{t\eta^2(d-1)}{2\sinh(t)}\left(1+4\xi\sinh^2(t/2)\right)\right] \ .
\end{equation}
So the approximate Green function is given by
\begin{eqnarray}
\label{Green-d}
G_E^{(d)}(r,r')&=&\frac1{2^{d+1/2}\pi^{d/2+1}}\frac{\Gamma(d/2)}
{(rr')^{\frac{d-1}2}}\int^\infty_\rho dt\frac1{\sqrt{\cosh t-\cosh\rho}}\times
\nonumber\\
&&\frac{\cosh(t/2)}{\sinh^{d-1}(t/2)}\left[1-\frac{t\eta^2(d-1)}{2\sinh(t)}
(1+4\sinh^2(t/2))\right] \ .
\end{eqnarray}

In his beautiful paper Christensen \cite{Christensen} has given a 
general expression for the Hadamard function for any dimensional spacetime, 
which is expressed in terms of the square of the geodesic distance 
$2\sigma(x,x')$. Moreover, there he has
called attention for the different behavior of this function when the
dimension of the spacetime is an even or odd number. In latter case there is 
no logarithmic terms in the expansion of the Hadamard. So because of this 
fact we shall develop, separately, the calculation of the renormalized vacuum 
expectation value of the square of the field operator, for $n=1+d$ odd and 
even.

Following \cite{Christensen}, below we write down the Hadamard function 
for the massless case when the dimension of the spacetime be an odd number. 
This functions is given by
\begin{equation}
G_H(x,x')=\frac{\Delta^{1/2}(x,x')}{2(2\pi)^{n/2}}\frac1{\sigma^{n/2-1}(x,x')}
\sum_{k=0}^{\frac{n-3}2}a_k(x,x')\sigma^k(x,x')\frac{\Gamma(n/2-k-1)}{2^k} \ ,
\end{equation}
where $\Delta(x,x')$, the Van Vleck-Moretti determinant and the factor
$a_k(x,x')$, for $k=0, \ 1,\ 2$, have been computed by many 
authors. See Refs. \cite{BS} and \cite{Christensen1}.

Now let us apply this formalism for the case $n=5$, i.e., $d=4$. The
Euclidean Green function in this case is
\begin{eqnarray}
\label{Green-App}
G_E^{(4)}(r,r')&=&\frac1{16\sqrt{2}\pi^3}\frac1{(rr')^{3/2}}\int^\infty_\rho
dt \frac1{\sqrt{\cosh t-\cos\rho}}\times
\nonumber\\
&&\frac{\cosh(t/2)}{\sinh^3(t/2)}\left[1-\frac{3t\eta^2}{2\sinh(t)}(1+
4\xi\sinh^2(t/2))\right] \ .
\end{eqnarray}

Because we need to evaluate the Hadamard function in the coincidence
limit we can write this function exhibiting only its divergent contributions
as shown below:
\begin{equation}
\label{Hadamard}
G_H(x,x')=\frac1{16\sqrt{2}\pi^2}\frac1{\sigma^{3/2}(x,x')}\left[1+
(1/6-\xi)R(x)\sigma(x,x')\right] \ ,
\end{equation}
where we have substituted the explicit expression for $\Delta$,  $a_0$ and 
$a_1$ in the coincidence limit. The scalar curvature in this five-dimensional
spacetime is $R=6\eta^2/r^2$. The radial one-half of the geodesic distance,
$\sigma(x,x')=(1/2\alpha^2)(r-r')^2$, in our approximation is equal to
$\sigma\approx(1/2)(r-r')^2(1+\eta^2+...)$. Now substituting
(\ref{Green-App}) and (\ref{Hadamard}) into (\ref{Phi2}) we get after a long
calculation
\begin{equation}
\langle\Phi^2(x)\rangle_{Ren.}=\frac{3\eta^2}{64\pi r^3}\left(\xi-
3/16\right) \ .
\end{equation}
We can see that for the conformal coupling in five dimensional spacetime, 
$\xi=3/16$, the above expectation vanishes, i.e., the renormalized vacuum 
expectation value of the operator $\Phi^2(x)$ is zero up to the first order in
$\eta^2$.

The six dimensional case will be analysed now. This case, together
with the four dimensional one studied in \cite{Mazzitelli}, exhibit
explicitly the ambiguity in the renormalization procedure given by a mass
subtraction point $\mu$. (Moretti in \cite{Moretti}, has used the local 
$\zeta$-function renormalization technique to show, by a general 
argument, that the scale ambiguity is present in the calculation of vacuum 
expectation value of the square of scalar field operator in a curved 
spacetime of dimension even.)  

The Hadamard function in even dimensions is explicitly written in the paper
by Christensen \cite{Christensen}, so we shall not reproduce it here. The 
singular behavior of the Hadamard function in the six dimensional 
spacetime is
\begin{equation}
\label{Hadamard1}
G_H(x,x')=\frac{\Delta^{1/2}(x,x')}{16\pi^3}\left[\frac{a_0(x,x')}
{\sigma^2(x,x')}+
\frac{a_1(x,x')}{2\sigma(x,x')}-\frac{a_2(x,x')}4\ln\left(\frac{\mu^2
\sigma(x,x')}2\right)\right] \ .
\end{equation}
Substituting the expressions for the factors $a_k$, we get, up to the first 
order in the parameter $\eta^2$, the following result: 
\begin{equation}
G_H(r,r')=\frac1{2\pi^3}\left[\frac{(1-2\eta^2)}{(r-r')^4}+
\frac{(1-6\xi)\eta^2}{2r^2(r-r')^2}-\frac{\eta^2}{4r^4}(\xi-1/5)
\ln\left(\frac{\mu^2(r-r')^2}4\right)\right] \ .
\end{equation}

Now taking $d=5$ in (\ref{Green-d}) and substituting the result, together
with the above equation, into (\ref{Phi2}) we get after some calculation
the following expression:
\begin{equation}
\langle\Phi^2(x)\rangle_{Ren.}=-\frac{\eta^2}{96\pi^3r^4}\left(\frac{47}{25}-
10\xi\right)+\frac{\eta^2}{8\pi^3r^4}\left(\xi-1/5\right)\ln(\mu r) \ .
\end{equation}
We can see that for the conformal coupling in six dimension, $\xi=1/5$, there
is no ambiguity, the logarithmic contribution disappears and we get
$\langle\Phi^2(x)\rangle_{Ren.}=\eta^2/800\pi^3r^4$. (In order to obtain the 
above result we expressed the logarithmic term which is present in 
(\ref{Hadamard1}) in terms of the $Q_0(\overline{u})$, $\overline{u}$ being 
$\frac{r^2+r'^2}{2rr'}$ and write this Legendre function in its integral 
representation (\ref{Legendre})).

The next section is devoted to the analysis of the vacuum expectation
value of the energy-momentum tensor in five and six dimensional spacetime.

\section{Vacuum Expectation Value of the Energy-Momentum Tensor}

In this paper we are working with a massless scalar quantum field theory
in the metric spacetime defined by (\ref{Metric}), which does not present
any dimensional parameter. Moreover we are adopting the natural system units 
where $\hbar=c=1$, so because of these reasons the physical quantities 
calculated in this model can only depend on the radial coordinate $r$ or 
on the renormalization mass scale $\mu$. By dimensional point we could 
expect that $\langle\Phi^2(x)\rangle_{Ren.}$ be proportional to $1/r^{n-2}$ 
and  $\langle T_{\mu,\nu}(x)\rangle_{Ren.}$ proportional to $1/r^n$. The 
factor of proportionality should be given in terms of the parameter $\eta^2$ 
and the non-minimal coupling $\xi$. As to the square of the scalar field,
this calculation has been done in this paper for the cases where the 
dimension of the spacetime is $5$ and $6$ up to the first order in $\eta^2$. 
In this section we want to analyse the vacuum expectation value of the 
energy-momentum tensor. We start considering the five dimensional case.

The renormalized vacuum expectation value of the energy-momentum tensor
in five dimensions does not depend on the subtraction mass parameter; so there
is no logarithmic term and it can be written in a general form by
\begin{equation}
\langle T_\mu^\nu(x)\rangle_{Ren.}=\frac{A_\mu^\nu(\xi,\eta^2)}{r^5} \ .
\end{equation}
Because there is no trace anomaly in odd dimension, for $\xi=3/16$, 
we can write
\begin{equation}
\langle T_\mu^\mu(x)\rangle_{Ren.}=0 \ ,
\end{equation}
so $A_\mu^\mu=0$. On the other hand, by symmetry of this spacetime, the above 
vacuum expectation value should be diagonal. Moreover the conservation condition
\begin{equation}
\nabla_\nu\langle T_\mu^\nu(x)\rangle_{Ren.}=0 \ ,
\end{equation} 
imposes additional restrictions on the components of the tensor $A_\mu^\nu$. So 
under these conditions we have:
\begin{equation}
A_0^0=A_1^1 ,\  A^2_2=A_3^3=A_4^4 \ .
\end{equation}
Using these relations and the traceless condition we can express all nonzero 
components of $A_\mu^\nu$ in terms of one of them, let us choose $A_0^0$, 
so we can write
\begin{equation}
A_\mu^\nu=A_0^0\ diag\ (\ 1,\ 1,\ -2/3,\ -2/3,\ -2/3\ ) \ .
\end{equation}
The explicit calculation of the component $A_0^0$ involves an extensive 
calculation which we shall not do here.

The vacuum expectation value of the energy-momentum tensor in six 
dimensions requires more details. By the expression obtained for the
vacuum expectation value of the square of the field,
it is possible to infer that there exists a logarithmic contribution
to this tensor. Moreover by the trace anomaly \cite{Christensen} we
have
\begin{equation}
\langle T_\mu^\mu(x)\rangle_{Ren.}=\frac1{64\pi^3}a_3(x) \ .
\end{equation}
So we can conclude that the general expression for this object is
\begin{equation}
\langle T_\mu^\nu(x)\rangle_{Ren.}=\frac1{64\pi^3r^6}
\left[A_\mu^\nu(\eta^2,\xi)+B_\mu^\nu(\eta^2,\xi)\ln(\mu r)\right] \ ,
\end{equation}
where $A_\mu^\nu$ in principal are arbitrary numbers. Because the
cutoff factor $\mu$ is completely arbitrary, there is an ambiguity
in the definition of this renormalized vacuum expectation value. 
Moreover the change in this quantity under the change of the 
renormalization scale is given in terms of the tensor $B_\mu^\nu$ as shown
below:
\begin{equation}
\langle T_\mu^\nu(x)\rangle_{Ren.}(\mu)-\langle T_\mu^\nu(x)\rangle_{Ren.}
(\mu')=\frac1{64\pi^3r^6}B_\mu^\nu(\eta^2,\xi)\ln(\mu/\mu') \ .
\end{equation}
In Ref. \cite{Christensen}, Christensen pointed out that the difference 
between them is given in terms of the effective action which depends
on the logarithmic terms whose final expression, in arbitrary even
dimension, is
\begin{equation}
\langle T_\mu^\nu(x)\rangle_{Ren.}(\mu)-\langle T_\mu^\nu(x)\rangle_{Ren.}
(\mu')=\frac1{(4\pi)^{n/2}}\frac1{\sqrt{g}}\frac\delta{\delta g^{\mu\nu}}
\int d^nx\sqrt{g}a_{n/2}(x)\ln(\mu/\mu') \ .
\end{equation}
In our six dimensional case we need the factor $a_3(x)$. The explicit
expression for this factor can be found in the paper by Gilkey \cite{Gilkey}
and in a more systematic form in the paper by Jack and Parker 
\cite{JP}, for a scalar second order differential operator $D^2+X$, 
$D_\mu$ being the covariant derivative including gauge field and $X$ an 
arbitrary scalar function. This expression involves $46$ terms and
we shall not repeat it here in a complete form. The reason is that our 
calculation has been developed up to the first order in the parameter 
$\eta^2$ and only the quadratic terms in Riemann and Ricci tensors, and 
in the scalar curvature are 
relevant for us. This reduces to $12$ the number of terms which will be 
considered. Discarding the gauge fields and taking $X=-\xi R$ we get:
\begin{eqnarray}
\overline{a}_3(x)&=&\frac16\left(\frac16-\xi\right)\left(\frac15-\xi\right)
R\Box R+\frac{\xi^2}{12}R^{;\mu} R_{;\mu}+\frac\xi{90}R^{\mu\nu}
R_{;\mu\nu}-\frac{\xi}{36}R^{;\mu} R_{;\mu}
\nonumber\\
&&-\frac1{7!}\left[28R\Box R +17R_{;\mu} R^{;\mu}-2R_{\mu\nu;\rho}
R^{\mu\nu;\rho}-4R_{\mu\nu;\rho}R^{\mu\rho;\nu}\right.+
\nonumber\\
&&9R_{\mu\nu\rho\sigma;\gamma}
R^{\mu\nu\rho\sigma;\gamma}-8R_{\mu\nu}\Box R^{\mu\nu}+24R_{\mu\nu}
R^{\mu\rho;\nu}\ _\rho+
\nonumber\\
&&\left.12R_{\mu\nu\rho\sigma}\Box R^{\mu\nu\rho\sigma}
\right]+O(R^3) \ .
\end{eqnarray}
This expression is of sixth order derivative on the metric tensor. Our 
next step is to take the functional derivative of $\overline{a_3}(x)$.
Using the expressions for the functional derivative of the 
Riemann and Ricci tensor, together with the scalar curvature \cite{BF1},
we obtain after a long calculation the following expression for the
tensor $B_\mu^\nu$:
\begin{eqnarray}
B_\mu^\nu(\eta^2,\xi)&=&\frac{r^6}6\left[-\delta_\mu^\nu\Box^2 R\left(
\xi^2-\frac\xi3+\frac{23}{840}\right)+\frac1{40}\Box^2 R_\mu^\nu+\right.
\nonumber\\
&&\left.\nabla^\nu\nabla_\mu\Box R\left(\xi^2-\frac\xi3+\frac1{42}\right)
\right]+O(R^2) \ .
\end{eqnarray}
Moreover, developing all the terms which appear in the above equation we obtain
after some calculations 
\begin{eqnarray}
B_\mu^\nu(\eta^2,\xi)&=&\frac{\eta^2}{225}\ diag\ (\ 6, 6, -3, -3, -3,
 -3\ )+
\nonumber\\
&&16\eta^2(\xi-1/5)(\xi-2/15)\ diag\ (\ 1, -4, 2, 2, 2, 2\ ) \ .
\end{eqnarray}

As in the last case it is possible to make some restrictions on the
tensor $A_\mu^\nu$. Again because of the spherical symmetry of the
problem we can infer that this tensor should be diagonal. Moreover
the renormalized vacuum expectation value of the energy-momentum 
tensor must be conserved, i.e.:
\begin{equation}
\langle T_\mu^\nu(x)\rangle_{Ren.;\nu}=0 \ .
\end{equation} 
From these six equations and defining the 
variable $T=\ 64\pi^3r^6\langle T_\mu^\mu(x)\rangle_{Ren.}$ we obtain
\begin{equation}
A_0^0=T+A_1^1-B_1^1+(B_1^1-B_0^0)\ln(\mu r)
\end{equation}
and
\begin{equation}
A_2^2=A_3^3=A_4^4=A_5^5=\frac{B_1^1}4-\frac{A_1^1}2-\left(B_2^2+
\frac{B_1^1}2\right)\ln(\mu r) \ .
\end{equation}
When the non-minimal coupling $\xi$ coincides with the conformal one,
$1/5$, we get
\begin{equation}
A_0^0=T+A_1^1-B_1^1 \ ,
\end{equation}
\begin{equation}
A_2^2=A_3^3=A_4^4=A_5^5=\frac{B_1^1}4-\frac{A_1^1}2 \ ,
\end{equation}
with
\begin{equation}
T=r^6 a_3(x)=-\frac1{4200}\Box^2 R +O(R^2)=\frac8{350}\frac{\eta^2}{r^6}
+O(\eta^4) \ .
\end{equation}
Again the complete evaluation of $\langle T_\mu^\nu(x)\rangle_{Ren.}$
requires the knowledge of at least one component of the tensor 
$A_\mu^\nu$, say $A_1^1$. However we do not attempt to do this 
straightforward and long calculation here.

\section{Concluding Remarks}

In this paper we have found, in a formal expression for the Euclidean
scalar Green function associated with a massless field in higher
dimensional global monopole spacetime defined by (\ref{Metric}), i. e., 
spacetime where the dimension is bigger than four and presents a solid
angle deficit. The expression for
this function is given in terms of an infinite sum of products of Legendre
functions with Gegenbauer polynomials or hyperspherical harmonics.
Having this Green function in our hands we can use it to calculate the
vacuum expectation value of some physically relevant operators, as 
the square of the field and the energy-momentum tensor. We have applied 
this formalism to calculate $\langle\Phi^2(x)\rangle$ and $\langle 
T_{\mu\nu}(x)\rangle$ in five and six dimensions. However these 
calculations become effective only for the case when the parameter $\alpha$,
associated with the solid angle deficit, is close to unity. In this
case we can expand the Green function in powers of the parameter $\eta^2=
1-\alpha^2$, and obtain closed results for these two quantities. 

As it was mentioned the vacuum expectation values for these quantities
are divergent and these divergences are consequence of the evaluation of 
the two points  Green  function in the coincidence limit. In order to 
obtain finite and well defined results we adopted the point-splitting 
renormalization procedure and  eliminate all divergences subtracting from 
the Green function the Hadamard one. An interesting result of our calculation
was that the renormalized vacuum expectation value of the square of the
field in the five dimensional global monopole spacetime vanishes when
we take for the non-minimal coupling constant $\xi$ the conformal
value $3/16$. A similar calculation in the six dimensional case shows that
a finite contribution remains for $\xi=1/5$ which is independent of
the mass cutoff parameter.

Moreover, in six dimensional case, there appears in the renormalized 
vacuum expectation value of the energy-momentum tensor, the function
$a_3(x)$. This function, according to \cite{Birrel} on pp. $159$, is
associated with a purely geometric (divergent) Lagrangian that should
renormalize the modified classical Einstein one. When similar terms are 
inserted into the gravitational action, the field equation is modified
by the presence of order six  terms proportional to:
\begin{equation}
c_1 g_{\mu\nu}\Box^2 R+c_2 \Box^2 R_{\mu\nu}+c_3 \nabla_\mu\nabla_\nu \Box R +
O(R^2) \ .
\end{equation}

\newpage
{\bf{Acknowledgments}}
\\       \\
We would like to thank Conselho Nacional de Desenvolvimento Cient\'\i fico e 
Tecnol\'ogico (CNPq) for partial financial support.

\section{Appendix}

In this appendix we present a non-trivial extension of our Green function
(\ref{Green}) for the case where $d=2$. Although this Green function 
constructed for $d\geq3$, presents a pole at $d=2$, the Gegenbauer polynomial 
is also not defined when its order is zero. However it is possible to get 
such expression if we  admit that the (\ref{Green}) is a function of a 
continuous parameter $d$. According to \cite{G}, it is possible to obtain a 
well defined limit for the ratio of the Gegenbauer polynomial by its order 
when it goes to zero. This limit is given by
\begin{equation}
\lim_{d\to2}\frac1{d-2}C_n^{\frac{d-2}2}(x)=\frac{T_n(x)}n \ ,
\nonumber\\
\end{equation}
where $T_n(x)$ is the Chebychev polynomials type $I$. Moreover in the limit 
$d=2$,
$\nu_n=|n|/\alpha$. However we have to be careful when we try to substitute 
the above limit into (\ref{Green}). The definition of the Chebychev
polynomials type $I$ by its generating function reproduces explicitly the 
polynomials with order $n\geq1$ in a recurrence equation separated from
the $T_0(x)$, see \cite{G}. Taking into account this fact we have to 
consider a multiplicity factor in the summation. Finally we arrive at
the following Green function is:
\begin{equation}
G^{(2)}(x,x')=\frac1{2\pi^2}\frac1{\sqrt{rr'}}\sum_{n=0}^\infty 
Q_{n/\alpha-1/2}T_n(\cos\gamma)\epsilon(n) \ ,
\end{equation}
with $\epsilon(0)=1$ and $\epsilon(n>0)=2$. However $T_n(\cos\gamma)=
\cos(n\gamma)$, see again \cite{G} on pp. 631. Finally substituting the 
integral representation for the Legendre function (\ref{Legendre}), we obtain
\begin{equation}
G^{(2)}(x,x')=\frac1{2\pi^2}\frac1{\sqrt{2rr'}}\int_\rho^\infty dt
\frac1{\sqrt{\cosh t-\cosh\rho}}\frac{\sinh(t/\alpha)}{\cosh(t/\alpha)-
\cos(\gamma)} \ .
\end{equation}
This equation is equivalent with the Euclidean Green function for the 
three dimensional conical space given in Eq. $(2.19)$ of \cite{Smith}. We 
can see that changing in a compatible way the coordinates $r$ by $r/\alpha$, 
$\gamma$ by $\gamma/\alpha$ and $\tau$ by $z$.


\begin{thebibliography}{100}
\bibitem{BF1} Reference [7], pp. 130-134. In this reference the 
convention for the Riemann tensor is different from the used here 
which makes its sign opposite to ours.
\end{thebibliography}
\end{document}